\documentclass[twocolumn,showpacs,preprintnumbers,amsmath,amssymb,prl,superscriptaddress]{revtex4}

\usepackage{amsmath,amssymb,amsthm,color}
\usepackage{graphicx}
\usepackage{dcolumn}

\newcommand {\beq}{\begin{eqnarray}}
\newcommand {\eeq}{\end{eqnarray}}

\begin{document}

\preprint{IPMU11-0172}

\title{On $\epsilon$-conjecture in $a$-theorem}

\author{Yu Nakayama}

\affiliation{Institute for the Physics and Mathematics of the Universe,  \\ Todai Institutes for Advanced Study,
University of Tokyo, \\ 
5-1-5 Kashiwanoha, Kashiwa, Chiba 277-8583, Japan}

\affiliation{California Institute of Technology, 452-48, Pasadena, California 91125, USA}


\begin{abstract}
The derivation of the $a$-theorem recently proposed by Komargodski and  Schwimmer relies on the $\epsilon$-conjecture that demands decoupling of dilaton from the rest of the infrared theory. We point out that the decoupling, if true, provides a strong evidence for the equivalence between scale invariance and conformal invariance in four dimension. Thus, a complete proof of the $a$-theorem along the line of their argument in the most generic scenario would establish the equivalence between scale invariance and conformal invariance, which is another long-standing conjecture in four-dimensional quantum field theories. 
\end{abstract}

\maketitle

\section{1. Introduction}
Recently  Komargodski and  Schwimmer \cite{Komargodski:2011vj} proposed an ingenious and illuminating derivation of the conjectured $a$-theorem \cite{Cardy:1988cwa}\cite{Jack:1990eb} --- there exists a quantity called $a$ that monotonically decreases along the renormalization group flow in four dimensional quantum field theories. They considered the low-energy effective action of the ``dilaton" that compensates the violation of the Weyl anomaly through the renormalization group flow from the  ultraviolet fixed point to the infrared fixed point, and argued that positivity constraint of the 2-2 scattering amplitude of the dilatons, which depends on the difference of the central charge $a_{\mathrm{UV}}- a_{\mathrm{IR}}$ gives the desired $a$-theorem.

In this paper, we point out that their argument relies on the $\epsilon$-conjecture that demands decoupling of dilaton from the rest of the infrared theory. It turns out that the decoupling, if true, provides a strong evidence for the equivalence between scale invariance and conformal invariance in four dimension \footnote{As we will discuss in section 2, when both infrared and ultraviolet fixed points are conformal, it is not difficult to argue that the dilaton will effectively decouple, so our discussion is by no means intended to invalidate their derivation of the $a$-theorem when the renormalization group flow is between two different conformal field theories. We would like to thank  Z.~Komargodski and S.~Rychkov for comments.}.

Whether scale invariant unitary relativistic quantum field theories in four dimension actually show conformal invariance is another long-standing problem in theoretical physics. In two dimension, the equivalence was proved \cite{Zamolodchikov:1986gt}\cite{Polchinski:1987dy}\cite{Mack1}, and in higher dimension with $d>4$, there is at least one explicit counterexample \cite{Jackiw:2011vz}\cite{ElShowk:2011gz}. A search for counterexamples in $d = 4-\epsilon$ has been discussed \cite{Dorigoni:2009ra}\cite{Fortin:2011ks}\cite{Fortin:2011sz}\cite{Nakayama:2011tk}. We also have a gravitational argument supporting the equivalence \cite{Nakayama:2009qu}\cite{Nakayama:2009fe}\cite{Nakayama:2010wx}.

The intimate connection between the $a$-theorem and the equivalence between scale invariance and conformal invariance is not unexpected. After all, the proofs of the corresponding statements in two-dimension go in complete parallel with each other: they come from the same set of assumptions with almost identical arguments \cite{Zamolodchikov:1986gt}\cite{Polchinski:1987dy}.  In addition, in holographic discussions that apply in arbitrary dimensions,  both of them are proved under the assumption of the null-energy condition \cite{Nakayama:2010wx}\cite{Girardello:1998pd}\cite{Freedman:1999gp}\cite{Myers:2010xs}\cite{Myers:2010tj}. In this paper, we would like to argue that a complete proof of the $a$-theorem along the line of the argument by  \cite{Komargodski:2011vj} in the most generic scenario would establish the equivalence between scale invariance and conformal invariance, revealing the intimate connection  in four dimension as well.

The rest of the paper is organized as follows. In section 2, we revisit the derivation of the $a$-theorem in the case where the renormalization group flow is induced by Higgsing, clarifying some assumptions made in the original paper \cite{Komargodski:2011vj}. In particular, we show that the decoupling of the dilaton is deeply related to the equivalence between scale invariance and conformal invariance in four dimension. 
In section 3, we extend the analysis in the case where the renormalization group flow is generated by relevant deformations. In section 4, we discuss further directions and conclude.

\section{2. $\epsilon$-conjecture in Higgsing}
We begin with the case where the renormalization group flow is induced by the so-called Higgsing. Suppose we start with an ultraviolet conformal field theory which has a moduli space of vacua. At a different point in the moduli space, the conformal invariance is spontaneously broken and the corresponding Nambu-Goldstone boson ($=$ dilaton $\tau$) appears. In addition, there may exist an infrared quantum field theory at each points of the moduli space. The question we want to ask is how the difference of the ultraviolet central charge $a_{{\mathrm{UV}}}$ and the infrared central charge $a_{{\mathrm{IR}}}$ behaves \footnote{We recall that the ``conformal anomaly" $a$ or $c$ can be defined for scale invariant but non-conformal field theories as the $c$-number violation of the tracelessness of the energy-momentum tensor: $\left\langle T^{\mu}_{\ \mu} \right\rangle = \left \langle D^\mu J_{\mu} \right \rangle  + a (\mathrm{Euler}) + c (\mathrm{Weyl})^2 + bR^2$. For instance, the ``conformal anomaly" for the linearized graviton, which is scale invariant but {\it not} conformal invariant \cite{Dorigoni:2009ra}, was defined and computed (see e.g. \cite{Birrell:1982ix} and references therein). Note that for non-conformal field theory, $b$ may not vanish. This will never cause a problem in our following discussion because we always embed the scale invariant field theory into a conformal field theory and $b$ will never appear there. Similarly, we should understand that the ``conformal anomaly" $a$ of the scale invariant but non-conformal field theory we study is that of the conformally embedded theory (minus $1$ from the arbitrary weakly coupled dilaton). We would like to thank Z.~Komargodski for related discussions.}.

In \cite{Komargodski:2011vj}, it was argued that the low-energy effective action of the setup after Higgsing is given by 
\begin{align}
S_{\mathrm{eff}} &= S_{\mathrm{QFT}_{\mathrm{IR}}} + f^2\int d^4x e^{-2\tau} (\partial_\mu \tau)^2 + \cr
&+ (a_{\mathrm{UV}} - a_{\mathrm{IR}}) \int d^4x  \left(2(\partial_\mu \tau)^4 -4(\partial_\mu \tau)^2 \Box \tau \right)   \label{KSeff}
\end{align}
with omitted higher derivative terms that are irrelevant for us. 
Here, $S_{\mathrm{QFT}_{\mathrm{IR}}}$ is the effective action of the infrared quantum field theory. $f$ is the decay constant of the dilaton, and most importantly, the coefficient in front of the four-dilaton derivative self-interaction is fixed by the anomaly matching of the Weyl anomaly. Then, the analytic properties of the 2-2 scattering of dilatons in the forward limit $A(s) \sim 2(a_{\mathrm{UV}} - a_{\mathrm{IR}}) \frac{s^2}{f^4}$ where $s$ is the Mandelstam $s$-variable,  yields a strong constraint on the sign of the four-dilaton derivative self-interaction \cite{Adams:2006sv}\cite{Komargodski:2011vj}, resulting in the desired inequality $a_{{\mathrm{UV}}} > a_{{\mathrm{IR}}}$. This is closely related to the absence of the superluminal propagation \cite{Adams:2006sv}.

We note that the effective action \eqref{KSeff} is by no means the most generic action that preserves the symmetry. In particular, the crucial assumption made in the original paper \cite{Komargodski:2011vj}, which we would like to call ``$\epsilon$-conjecture" is that the dilaton does not interact with the remaining infrared quantum field theory, or more precisely the interaction is sufficiently suppressed so that it would not affect their analysis on the 2-2 scattering process of dilatons. As we will see, the introduction of the interaction between the dilaton and the infrared field theory may spoil the above simple argument of the positivity.

Before we discuss such a possibility, we would like to point out that if the $\epsilon$-conjecture, or the decoupling of the dilaton is true, the infrared quantum field theory under consideration is not only scale invariant but also conformal invariant. Hence, the derivation of the $a$-theorem is directly related to the claim of the equivalence between scale invariance and conformal invariance. To see this, we note that the ultraviolet conformal field theory is Weyl invariant by assumption, so the total energy-momentum tensor must be traceless (up on improvement) as an operator identity:
\begin{align}
T^{\mu}_{\ \mu (\mathrm{tot})} = 0 \ . \label{tot}
\end{align}
This means that the {\it sum} of the energy-momentum tensor in the infrared quantum field theory and that of the dilaton must vanish in the far infrared under the assumption of the decoupling. Since the dilaton part is Weyl (conformal) invariant by itself, the energy-momentum tensor of the decoupled infrared quantum field theory must be traceless:
\begin{align}
T^{\mu}_{\ \mu (\mathrm{QFT}_{\mathrm{IR}})} = 0 \ .
\end{align}
In other words, the dilaton part of the action is fixed by requiring the quantum Weyl invariance, so the decoupled infrared part of the action must be Weyl invariant as well to preserve the total Weyl invariance that is spontaneously broken \footnote{A technical comment is in order. We here assume that the energy-momentum tensor of the infrared theory is given by the restriction of the {\it same} energy-momentum tensor of the ultraviolet theory. Since non-trivial accidental symmetry may appear in the infrared and the energy-momentum tensor always has an ambiguity of the improvement, this is not at all trivial. However, we claim that this assumption is a part of the {\it definition} of Higgsing. Otherwise, we cannot tell which is ultraviolet and which is infrared by moving the moduli space of vacua. Indeed, if we started with the ``Higgsed" theory and move along the moduli space to reach the ``unHiggsed" theory, then the energy-momentum tensor of the ``unHiggsed" theory would not be obtained by the restriction of that of the ``Higgsed" theory because of the appearance of the new massless degrees of freedom.}.

This shows that the infrared quantum field theory under the assumption of the decoupled dilaton is not only scale invariant but also conformal invariant because the energy momentum tensor of the decoupled infrared field theory is traceless. This is a desirable result because the repeated Higgsing will never produce scale invariant but not conformal invariant field theories.

The question we should ask, however, is what happens if the infrared quantum field theory is only scale invariant but not conformal invariant. It is not difficult to see that, in this circumstance, the decoupling {\it must} be violated. Due to the assumed conformal invariance of the ultraviolet theory (which we would like to argue momentarily how to relax), the total energy momentum tensor is still traceless as in \eqref{tot}. The scale but non-conformal invariance of the infrared quantum field theory means that the trace of its energy-momentum tensor is non-zero and is given by the divergence of the Virial current $J^{\mu}$:
\begin{align}
T^{\mu}_{\ \mu (\mathrm{QFT}_{\mathrm{IR}})} = \partial^\mu J_\mu \ . \label{Viri}
\end{align}
This means that the dilaton cannot decouple from the infrared quantum field theory because otherwise the dilaton sector is Weyl invariant by itself and we cannot assure the total Weyl invariance.
The way to circumvent this inconsistency is to introduce a coupling between the  dilaton and the infrared scale invariant field theory. Indeed, the violation of the Weyl invariance from \eqref{Viri} should be compensated by the coupling
\begin{align}
\delta S = \int d^4x  J^\mu \partial_\mu \tau\ \label{comp}
\end{align}
to ensure the total Weyl invariance by noting $\tau \to \tau + \sigma$ under the Weyl transformation \footnote{We may also introduce additional curvature couplings that will compensate a ``removable" part of the Virial current by improving the energy-momentum tensor. The introduction of the dilaton coupling is imperative only when the energy-momentum tensor cannot be improved.}.

For illustrative purposes, let us show an example. Of course, there is no known scale invariant but non-conformal unitary quantum field theory in four-dimension, so the example here is non-unitary. The model is essentially four-dimensional version of the one studied by Riva and Cardy \cite{Riva:2005gd} (i.e. the most generic massless vector field theory). Suppose the infrared quantum field theory that we have discussed is given by the action
\begin{align}
S = \int d^4 x \left(\frac{1}{4}F_{\mu\nu} F^{\mu\nu} + \epsilon (\partial_\mu A^\mu)^2 \right)\ ,
\end{align} 
where $F_{\mu\nu} = \partial_\mu A_\nu - \partial_\nu A_\mu$. The Weyl invariance is broken with non-zero $\epsilon$, and the trace of the energy-momentum tensor is given by
\begin{align}
T^{\mu}_{\ \mu} = \epsilon( 2 A_\mu (\partial^\mu \partial^\nu) A_\nu -2 (\partial^\mu A_\mu)^2) \ .
\end{align}
The violation is due to the Christoffel symbol appearing in the covariant derivative of the divergence $(D_{\Gamma}^\mu A_\mu)$ when we put the theory on a curved background.

It is easy to recover the Weyl invariance by introducing the Weyl covariant derivative $D_W^\mu A_\mu = (\partial^\mu + \partial^\mu \tau  ) A_\mu$. If the action was given by Higgsing of the ultraviolet conformal field theory as we have discussed, the Weyl covariantization must have been automatic. Indeed, the replacement of the covariant derivative yields the first order interaction
\begin{align}
\delta S = \epsilon \int d^4 x \partial_\mu \tau A^{\mu} (\partial^\nu A_\nu) \ ,
\end{align}
between the dilaton and the vector field,
which is nothing but the compensating coupling between dilaton and the Virial current \eqref{comp} up to equations of motion. This illustrates how the total Weyl invariance demands the existence of the interaction between scale invariant but non-conformal infrared quantum field theory and the dilaton.

In summary, as the Weyl anomaly induced by the central charge determines the four-dilaton derivative self-interaction, the Virial current contributions to the trace of the energy-momentum tensor determines the coupling between the scale invariant but non-conformal infrared field theory and the dilaton as a low-energy theorem.
We have seen that the $\epsilon$-conjecture that states the decoupling of the dilaton in the infrared implies the equivalence between scale invariance and conformal invariance. 

A question arises whether the derivation of the $a$-theorem \cite{Komargodski:2011vj} goes through even though the dilaton would not decouple. Unfortunately, this is not the case. When the virial current four-point correlation function $\left \langle J J JJ \right \rangle$ does not vanish, which is generically expected, the 2-2 dilaton scattering in the forward limit obtains additional contribution of the order of $\frac{s^2}{f^4}$ from the exchange of the Virial current, which is comparable to the one coming from the four-dilaton derivative self-interactions proportional to $a_{\mathrm{UV}} - a_{\mathrm{IR}}$. Then it is impossible to argue the positivity of $a_{\mathrm{UV}} -a_{\mathrm{IR}}$ from the analytic structure of the 2-2 scattering amplitudes.

Note that when the Virial current is conserved, the leading derivative 2-2 scattering amplitude from the exchange of the Virial current vanishes. Similarly, when the Virial current is a derivative of a certain tensor operator i.e. $J^{\mu} = \partial^{\nu} \Lambda_{\ \nu}^{\mu}$, the leading amplitude vanishes due to the on-shell condition of the dilaton \footnote{A simple example is a sigma model whose target space is $\mathbf{R}^2$. In the polar coordinate, the Lagrangian is $(\partial_\mu R)^2 + R^2 (\partial_\mu \theta)^2$, and the dilaton $\tau = \log R $ does not look ``decoupled" from the angular variable field $\theta$ with non-zero interaction $\delta L = \partial_\mu \tau J^\mu$. The virial current $J_\mu = \theta \partial_\mu \theta$, however, is a total derivative, and the S-matrix vanishes as expected.}. This is all consistent with the ambiguity of the energy-momentum tensor and Virial current (see e.g. \cite{Polchinski:1987dy}), and it indicates that precisely when the theory is scale invariant but not conformal invariant, the derivation of the $a$-theorem proposed in \cite{Komargodski:2011vj} cannot be completed.

In this discussion, we have assumed that the starting ultraviolet theory is conformal invariant rather than merely scale invariant. In a more generic situation, we had to allow a scale invariant but non-conformal field theory as an ultraviolet theory. In order to complete the analysis in this case, we have to introduce a compensating dilaton in the ultraviolet theory with appropriate kinetic function by hand, and require the Weyl invariance of the total action. The total Weyl invariance is imperative so that the anomaly matching makes sense.
At the first order, we add 
\begin{align}
\delta S_{UV} = f^2 \int d^4x e^{-2\tau} (\partial_\mu \tau)^2 +  \int d^4x J^{\mu}_{\mathrm{UV}} \partial_\mu \tau + ...
\end{align}
with the ultraviolet Virial current $J_{\mathrm{UV}}^\mu$.
 The situation is very close to the one where the renormalization group flow is induced by adding the relevant perturbation, which we will discuss in the next section.
Note that at this point, the theory augmented by the compensating dilaton is conformal invariant, so the notion of central charge $a$ is free from the subtlety of the scale invariant but non-conformal field theory. We may use it as a definition of the central charge of the scale invariant but non-conformal field theory.

After the spontaneous breaking of the Weyl invariance, which is introduced by hand with the compensating dilaton, due to Higgsing, we would like to analyze the sign of $a_{\mathrm{UV}} - a_{\mathrm{IR}}$ by studying the 2-2 scattering of the Nambu-Goldstone dilatons. Again, if we can show that the Nambu-Goldstone dilaton decouples from the rest of the infrared degrees of freedom, the $a$-theorem immediately follows from the  analytic structure of the scattering amplitudes, but the dilaton decoupling is not at all obvious since even in the ultraviolet theory, it did not decouple (due to the scale but non-conformal nature of the ultraviolet theory).

\section{3. Compensating dilaton and $\epsilon$-conjecture}
In \cite{Komargodski:2011vj}, a generalization of the argument where the renormalization group flow is induced by adding the relevant perturbation to the ultraviolet fixed point theory was discussed. Again, their argument relies on the decoupling of the dilaton among other subtleties that we will explain. 

The starting point is to introduce the compensating dilaton so that the relevant perturbation is uplifted to be Weyl invariant. This is achieved by augmenting the Weyl weight factor appearing in the deformation by the compensating dilaton $e^{\tau}$. We will also introduce the kinetic term for the compensating dilaton  with large decay constant by hand so that the dilaton will not change the dynamics. The first implicit assumption in \cite{Komargodski:2011vj}  is that the resulting Weyl invariant theory is ultraviolet completed, unitary and stable.

This assumption looks innocuous, but it is not obvious. For instance, let us consider the $(1+1)$ dimensional sin-Gordon theory, which is not conformal invariant. We may  dress the sin-Gordon potential $\sin(\alpha X)$ with the Liouville factor $e^{\beta \phi}$, which plays the role of the compensating dilaton, to make the theory look like conformal invariant. However, it is clear that the dressed sin-Liouville potential $e^{\beta \phi} \sin (\alpha X)$ is unbounded below, and the stability of the theory is questionable (at least from the classical viewpoint). At a very specific value of $\alpha$ and $\beta$, it is known that the theory {\it is} conformal invariant and stable (thanks to the existence of dual $SL(2,\mathbf{R})/U(1)$ coset model description \cite{FZZ}), but what happens in generic situations is unknown (see also \cite{Baseilhac:1998eq}
\cite{Nakayama:2006gt}). In the discussion by \cite{Komargodski:2011vj}  and here, the existence of the ultraviolet completed healthy conformal field theory obtained by the naive conformal compensation is assumed, but such an assumption can be highly non-trivial.

The second step is to study the 2-2 scattering amplitude of the compensating dilatons from the low-energy effective field theory viewpoint in the far infrared. Even though we assume that the introduction of the compensating dilaton gives a sensible ultraviolet completed theory, we still have to deal with the analogue of the $\epsilon$-conjecture discussed in the previous section here. Again in \cite{Komargodski:2011vj}  it was argued that the low energy-effective action takes the following decoupled form:
\begin{align}
S_{\mathrm{eff}} &= S_{\mathrm{QFT}_{\mathrm{IR}}} + f^2\int d^4x e^{-2\tau} (\partial_\mu \tau)^2 + \cr
&+ (a_{\mathrm{UV}} - a_{\mathrm{IR}}) \int d^4x  \left(2(\partial_\mu \tau)^4 -4(\partial_\mu \tau)^2 \Box \tau \right)  
\end{align} 
As discussed in the previous section, this decoupling is not at all trivial. If this decoupling were true, then the infrared theory induced by the renormalization group flow of the relevant deformation would be conformal invariant. On the other hand, when the infrared theory is scale invariant but not conformal invariant, the decoupling is inconsistent with the Weyl invariance of the total theory. There we had to add the coupling
\begin{align}
\delta S = \int d^4x  J^\mu \partial_\mu \tau\ \  \label{vint}
\end{align}
to compensate the non-zero trace of the energy-momentum tensor of the infrared scale invariant but non-conformal field theory.

In \cite{Komargodski:2011vj}, the ``correction" of this sort was excluded because of the naive dimensional reason --- these must be suppressed by the factor of higher $\log e^{\tau}/M$. This is indeed true, and the introduction of the dilaton would not change the structure of the infrared quantum field theory for large $f$. However, the real question we should ask is whether this log term would or would not introduce the coupling between the infrared quantum field theory and the dilaton sector that is relevant for the 2-2 dilaton scattering. The leading contribution \eqref{vint}, which appears when the infrared theory is scale invariant but not conformal invariant is as effective as the $(\partial_\mu \tau)^4$ term governed by the central charge as we have discussed in section 2 and it spoils the positivity argument of the scattering amplitudes.

In summary, we have clarified two main assumptions, among others, in the derivation of the $a$-theorem proposed by \cite{Komargodski:2011vj} in the case where the renormalization group flow is induced by adding relevant deformations. The first assumption, the stability and unitarity of the Weyl compensated theory needs deeper understanding. The second assumption, the $\epsilon$-conjecture seems deeply related to the (non-)existence of scale invariant but non-conformal field theories in four-dimension.

\section{4. Discussions}

In this paper, we have studied the consequence of the $\epsilon$-conjecture that demands the decoupling of dilaton from the infrared quantum field theories. On  one hand, the $\epsilon$-conjecture gives the foundation of the derivation of the conjectured $a$-theorem. On the other hand, the $\epsilon$-conjecture provides a strong support for the equivalence between scale invariance and conformal invariance. Unfortunately, we do not give a proof of the $\epsilon$-conjecture in this paper except that we point out  it is known classically as well as perturbatively at one-loop order the conjecture is true. 

We note that there exists an explicit counterexample of the equivalence between scale invariance and conformal invariance in unitary relativistic field theories in higher dimension than four, so we suspect the decoupling of the dilaton may not be true there. 

Finally, we should point out that the possibility to find a counterexample of the $a$-theorem does not perish until we give a full proof of the $\epsilon$-conjecture. Given the discussions in this paper, a candidate counterexample, if any, would show scale invariance but non-conformal invariance. Discussions on the relation between (the stronger version of) the $a$-theorem and the (in)equivalence between scale invariance and conformal invariance can be found in \cite{Fortin:2011sz}\cite{Nakayama:2011tk}. In such cases, it is also logically possible that  the $a$-theorem is only true for the renormalization group flow between conformal invariant field theories. Another possibility is that it is not $a$ defined through the obstruction of the Weyl gauging from the $c$-number violation of the tracelessness but something else that shows the monotonically decreasing property in the case when the theory is not conformal but only scale invariant.
We hope we are going to settle these problems in the near future.

\section*{Acknowledgments}
The author thanks H.~Ooguri for discussions.
The work is supported by the 
World Premier International Research Center Initiative of MEXT of
Japan. 


\end{document}